text

# Lanthanum-Cerium Based Bulk Metallic Glasses with Superior Glass-Forming Ability

Ran Li, Shujie Pang, Tao Zhang*

Department of Materials Science and Engineering, Beijing University of Aeronautics and Astronautics, Beijing 100083, China

**Abstract**

A quinary $(La_{0.5}Ce_{0.5})_{65}Al_{10}(Co_{0.6}Cu_{0.4})_{25}$ alloy with superior glass-forming ability (GFA), identified by the formation of fully glassy rod of 32 mm in diameter by tilt-pour casting, was reported. By comparing with the GFA of quarternary $(La_{0.5}Ce_{0.5})_{65}Al_{10}TM_{25}$ and ternary $Ln_{65}Al_{10}TM_{25}$ alloys (Ln = La or Ce; TM = Co or Cu), we suggest that the strong frustration of crystallization by utilizing the coexistence of La-Ce and Co-Cu to complicate competing crystalline phases is helpful to construct BMG component with superior GFA.

* Corresponding author. Tel: +86-10-82314869; fax: +86-10-82314869.

E-mail: zhangtao@buaa.edu.cn (T. Zhang).

## Introduction

In recent years, rare earth based bulk metallic glasses (BMGs) were focused by increasing interests because of their novel physical properties, such as high glass-forming ability (GFA) [1-3], special magnetic properties [4-6], superplasticity and thermoplasticity in supercooled liquid region [7-8], high fracture strength and ductility [2,9], which bring these BMGs potential applications in the future. Since a La-Al-Ni BMG was successfully synthesized by copper mold casting in 1989, the composition exploration for new BMGs with higher GFA has attracted widely enthusiasm because the GFA of metallic glasses is significant for scientific research and further commercial application [10,11]. In following years, great successes have been achieved in many alloy systems, such as Zr-, Pd-, Y-, Mg-, Pt-based etc. [1,12-19]. However, the superior GFA with the formation of glassy rods above 30 mm in diameter was only found in Zr-, Pd- and Pt-based metallic glasses, and never discovered in rare earth based BMGs.

Recently, we reported that the GFA of ternary single-lanthanide based BMGs, Ln-Al-Co (Ln=La, Ce, Pr and Nd, respectively) with the glassy critical diameter, $d_c$ ($d_c$ means the maximal diameter size of fully glassy rod which can be produced for a certain alloy), of not more than 4 mm, can be evidently improved by substituting multi-lanthanide elements for single-lanthanide solvent element to form (Ce-La-Pr-Nd)-Al-Co BMG with the $d_c$ of 15 mm [2]. Furthermore, more “simple” glassy alloys in a quaternary (La-Ce)-Al-Co system with higher GFA indicated by the $d_c$ up to 25 mm was pinpointed [20]. The obvious improvement of GFA implies that the coexistence of similar lanthanide elements with similar atomic sizes and various valence electronic structures has significant positive effect to the GFA of BMGs. In addition, the positive effect of coexistence of similar elements on GFA can be also found out in other glassy systems, such as the coexistence of Fe and Co in (Fe-Co)-(Si-B)-Nb and (Fe-Co)-(Cr-Mo)-(C-B)-Y, Fe and Ni in (Ni-Fe)-(Si-B)-Nb, Pt and Pd in (Pd-Pt)-Cu-P as solvent elements [16,21-23], Cu and Ni in Zr-Al-(Ni-Cu), Pt-(Ni-Cu)-P, Pd-(Ni-Cu)-P and La-Al-(Ni-Cu) as solute elements [3,12,14,24], in which GFA can be increased remarkably by the coexistence of similar elements.

However, little experimental and theoretical analysis has been performed on this coexistence influence.

In this work, we developed a family of La-Ce based BMGs with superior GFA identified by the formation of fully glassy rods up to 32 mm through the equiatomic ratio or near equiatomic ratio coexistence of similar element pairs of La-Ce and Co-Cu. The glass formation, thermal stability and melting behavior of quinary $(La_{0.5}Ce_{0.5})_{65}Al_{10}(Co_{0.6}Cu_{0.4})_{25}$ alloy, quarternary $(La_{0.5}Ce_{0.5})_{65}Al_{10}Co_{25}$ and $(La_{0.5}Ce_{0.5})_{65}Al_{10}Cu_{25}$ alloys were evaluated. For comparison, the ternary $La_{65}Al_{10}Co_{25}$, $Ce_{65}Al_{10}Co_{25}$, $La_{65}Al_{10}Cu_{25}$ and $Ce_{65}Al_{10}Cu_{25}$ alloys were also studied. It indicates that the frustration of crystallization by utilizing the coexistence of similar element pairs to complicate the competing crystalline phases is beneficial to improve the GFA of the "designed" BMG component.

**Experimental Procedure**

Alloy ingots with nominal compositions were prepared by arc-melting the mixture of the rare earth metals (>99.5 mass%), Al (>99.99 mass%), Co (> 99.9 mass%) and Cu (> 99.9 mass%) in a highly pure argon atmosphere. For the smaller rod-shaped sample (≤ 12 mm in diameter), the ingot was remelted in a quartz tube and then injected into a copper mold at a highly pure argon atmosphere to produce glassy rods. For the larger rod-shaped sample (> 12 mm in diameter), the ingot was remelted in a quartz cup using a tilting induction furnace and then the molten alloy was poured into a copper mold in a highly pure argon atmosphere. Phase structure of the samples was examined by X-ray diffraction (XRD) using a Bruker AXS D8 X-ray diffractometry with Cu-K$\alpha$ radiation at a scanning rate of 1 degree/minute. Thermal stability of the glassy samples was investigated by a NETZSCH DSC 404 C Differential Scanning Calorimeter (DSC) at a heating rate of 0.33 K/s. Melting and solidification behaviors of these alloys were also characterized by DSC at a heating and cooling rate of 0.33 K/s.

**Results and Discussion**

According to the former research of the coexistence influence of similar elements on GFA of resulting alloys [2,20], we adopted two pairs of similar elements, La-Ce and Co-Cu, with the coexistence of equiatomic or near equiatomic ratio to construct the alloy components with high GFA. The fully glassy rod of 32 mm in diameter, successfully produced by tilt-pour casting, confirms the "designed" alloy of $(La_{0.5}Ce_{0.5})_{65}Al_{10}(Co_{0.6}Cu_{0.4})_{25}$ has the superior GFA. Outer shape and surface appearance of the as-cast rod of 32 mm in diameter is shown in Fig. 1. The surface of sample is smooth and lustrous. The XRD pattern of this as-cast rod is shown in Fig. 2. The smooth broad diffuse peak indicates no crystalline phase in this specimen. The glass formation of $(La_{0.5}Ce_{0.5})_{65}Al_{10}Co_{25}$ and $(La_{0.5}Ce_{0.5})_{65}Al_{10}Cu_{25}$ alloys was also evaluated by copper mold casting. The two alloys have high GFA identified by the formation of fully glassy rod of 12 mm and 8 mm in diameter, respectively. The XRD patterns of these two BMGs, shown in Fig. 2, confirm no detectable crystalline phase in these corresponding samples.

Figure 3 shows the DSC curves of the $(La_{0.5}Ce_{0.5})_{65}Al_{10}(Co_{0.6}Cu_{0.4})_{25}$, $(La_{0.5}Ce_{0.5})_{65}Al_{10}Co_{25}$ and $(La_{0.5}Ce_{0.5})_{65}Al_{10}Cu_{25}$ BMG samples with the same diameter of 2 mm during the heating and cooling procedures at a rate of 0.33 K/s. In order to investigate the coexistence effect of similar element pairs, La-Ce and Co-Cu, with similar atomic sizes and various valence electronic structures on glass formation, four ternary alloys, $La_{65}Al_{10}Co_{25}$, $Ce_{65}Al_{10}Co_{25}$, $La_{65}Al_{10}Cu_{25}$ and $Ce_{65}Al_{10}Cu_{25}$, were also fabricated for comparison. Figure 4 shows the DSC curves of these ternary BMGs at the same heating and cooling rates. The thermal parameters, such as glass transition temperature $T_g$, crystallization temperature $T_x$, supercooled liquid region $\Delta T_x$ ($\Delta T_x = T_x - T_g$), melting temperature $T_m$, $T_l$ (liquidus temperature determined by heating procedure), $T_{l'}$ (liquidus temperature determined by cooling procedure), $T_g/T_m$, reduced glass transition temperature $T_{rg}$ ($T_{rg} = T_g/T_l$ or $T_g/T_{l'}$), $\gamma$ ( $\gamma = T_x / (T_g + T_l)$) [25,26], $\Delta T_l$ (the nominal degree of supercooling, $\Delta T_l = T_l - T_{l'}$) and $d_c$, for all glassy specimens are shown in Table I. We can list the GFA for all alloys from poor one to high one according to the corresponding $d_c$ values as follows: $Ln_{65}Al_{10}Co_{25}$ ($d_c$ = 2 mm) < $Ln_{65}Al_{10}Cu_{25}$ ($d_c$ = 4 mm) < $(La_{0.5}Ce_{0.5})_{65}Al_{10}Cu_{25}$ ($d_c$ = 8 mm) <

$(La_{0.5}Ce_{0.5})_{65}Al_{10}Co_{25}$ ($d_c$ = 12 mm) < $(La_{0.5}Ce_{0.5})_{65}Al_{10}(Co_{0.6}Cu_{0.4})_{25}$ ($d_c$ = 32 mm), in which Ln represents La or Ce. We found out that when each similar element pair was induced into the alloy component to substitute for the former single element, the GFA of resulting alloys could be improved remarkably. Although it is reasonable that the $(La_{0.5}Ce_{0.5})_{65}Al_{10}(Co_{0.6}Cu_{0.4})_{25}$ alloy has the highest GFA because of its largest $T_g/T_{l'}$ (0.66) and $\Delta T_l$ (114 K) comparing with other alloys, the criteria of GFA, $T_{rg}$, $\gamma$, and $\Delta T_x$, which can give the good indication for high GFA in many former glassy alloy systems [11,25,26], can't give the approperiate explanations for the GFA of all glassy alloys in Table I.

It is well known that the key point for glass formation of metallic glasses with high GFA is the frustration of crystallization during the solidification. Some researchers demonstrated how to complicate competing crystalline phases and frustrate the process of crystallization in multi-component alloys [19,27,28]. We evaluated the difference of eutectic composition and intermetallic phases in phase diagrams of La-Co, Ce-Co, La-Cu and Ce-Cu. For La-Co and Ce-Co, both show stoichiometric intermetallic phases, $La_3Co$ and $Ce_{24}Co_{11}$, with low melting temperature on the lanthanide-rich side. The stable competing crystalline phases might be responsible for the poor GFA of $La_{65}Al_{10}Co_{25}$ and $Ce_{65}Al_{10}Co_{25}$ ($d_c$ = 2 mm). However, when we utilized the coexistence of La and Ce elements to substitute for single- La or Ce element in ternary Ln-Al-Co alloys, the competing crystalline phases were complicated and the process of crystallization was strongly frustrated because $La_3Co$ has the $Fe_3C$-type structure, while $Ce_{24}Co_{11}$ is a complex hexagonal phase with space group $P6_3mc$ [29]. Therefore, the GFA can be prominently improved in the $(La_{0.5}Ce_{0.5})_{65}Al_{10}Co_{25}$ resulting alloy ($d_c$ = 12 mm) comparing with single-lanthanide based ternary alloys. Considering La-Cu and Ce-Cu phase diagrams, there is little discrepancy between the two diagrams and both show the deep eutectics at the similar Cu concentration on the lanthanide-rich side. So it could be easily understood that the GFA of $Ln_{65}Al_{10}Cu_{25}$ ($d_c$ = 4 mm) is larger than that of $Ln_{65}Al_{10}Co_{25}$ ($d_c$ = 2 mm) according to the Turnbull's work [25]. For the situation of the coexistence of La and Ce in $(La_{0.5}Ce_{0.5})_{65}Al_{10}Cu_{25}$ alloy, we noticed that in the La-Cu and Ce-Cu phase diagrams the possible completing intermetallic phases, LaCu and CeCu, have the same FeB-type structure. It implies that there is no obvious contribution of complicating crystallization through the coexistence of La and Ce in $(La_{0.5}Ce_{0.5})_{65}Al_{10}Cu_{25}$ alloy as the situation of $(La_{0.5}Ce_{0.5})_{65}Al_{10}Co_{25}$ alloy, proved by the fact that the $d_c$ of $(La_{0.5}Ce_{0.5})_{65}Al_{10}Cu_{25}$ is smaller than that of $(La_{0.5}Ce_{0.5})_{65}Al_{10}Co_{25}$. However, the GFA of the $(La_{0.5}Ce_{0.5})_{65}Al_{10}Cu_{25}$ alloy ($d_c$ = 8 mm) can be still improved comparing with that of $Ln_{65}Al_{10}Cu_{25}$ alloys ($d_c$ = 4 mm), because the chemical disorder may be increased in the resulting alloy [27]. Furthermore, based on the above analysis, because of the obvious distinct difference of phase diagrams between La-Co/Ce-Co and La-Cu/Ce-Cu, the possible competing crystalline phases of $(La_{0.5}Ce_{0.5})_{65}Al_{10}(Co_{0.6}Cu_{0.4})_{25}$ alloy become more complex and the process of crystallization is stronger frustrated so as to obtain the superior GFA of this alloy with $d_c$ up to 32 mm. Therefore, we suggest that the BMG component with superior GFA can be constructed through the multistep complication of competing crystalline phases and strong frustration of crystallization using the similar element coexistence in resulting alloys.

**Conclusions**

In this paper, a family of La-Ce based BMGs with superior GFA was reported. The fully glassy rods of $(La_{0.5}Ce_{0.5})_{65}Al_{10}(Co_{0.6}Cu_{0.4})_{25}$, $(La_{0.5}Ce_{0.5})_{65}Al_{10}Co_{25}$ and $(La_{0.5}Ce_{0.5})_{65}Al_{10}Cu_{25}$ can be successfully produced up to 32 mm, 12 mm and 8 mm in diameter, respectively. Comparing with the GFA of ternary $Ln_{65}Al_{10}TM_{25}$ alloys (Ln = La or Ce; TM = Co or Cu), the largest reduced glass transition temperature ($T_g/T_l$=0.66) and nominal degree of supercooled ($\Delta T_l$ = 114 K) of $(La_{0.5}Ce_{0.5})_{65}Al_{10}(Co_{0.6}Cu_{0.4})_{25}$ alloy can give a good explanation of the highest GFA. Furthermore, we suggest that the similar element coexistence of La-Ce and Co-Cu is beneficial to complicate competing crystalline phases and frustrate the process of crystallization, which result in the superior GFA in $(La_{0.5}Ce_{0.5})_{65}Al_{10}(Co_{0.6}Cu_{0.4})_{25}$ alloy. The coexistence effect of similar elements is helpful for us to "design" the BMG component with superior GFA.

**Acknowledgements**

The authors would like to thank Prof. A. R. Yavari for many constructive suggestions improving the manuscript. This work was financially supported by National Nature Science Foundation of China (No. 50225103, No. 50471001 and No. 50631010).

**Figure Captions**

Fig. 1 The photo of as-cast 32 mm diameter glassy rod of $(La_{0.5}Ce_{0.5})_{65}Al_{10}(Co_{0.6}Cu_{0.4})_{25}$.

Fig. 2 XRD patterns of the as-cast rod of 32, 12 and 8 mm in diameter of $(La_{0.5}Ce_{0.5})_{65}Al_{10}(Co_{0.6}Cu_{0.4})_{25}$, $(La_{0.5}Ce_{0.5})_{65}Al_{10}Co_{25}$ and $(La_{0.5}Ce_{0.5})_{65}Al_{10}Cu_{25}$ alloys respectively.

Fig. 3 DSC curves of $(La_{0.5}Ce_{0.5})_{65}Al_{10}(Co_{0.6}Cu_{0.4})_{25}$, $(La_{0.5}Ce_{0.5})_{65}Al_{10}Co_{25}$ and $(La_{0.5}Ce_{0.5})_{65}Al_{10}Cu_{25}$ BMGs during melting and solidification processes.

Fig. 4 DSC curves of the ternary $Ln_{65}Al_{10}Co_{25}$ and $Ln_{65}Al_{10}Cu_{25}$ (Ln = La, Ce) BMGs during melting and solidification processes.

**Fig. 1 The photo of as-cast 32 mm diameter glassy rod**

Click here to download high resolution image

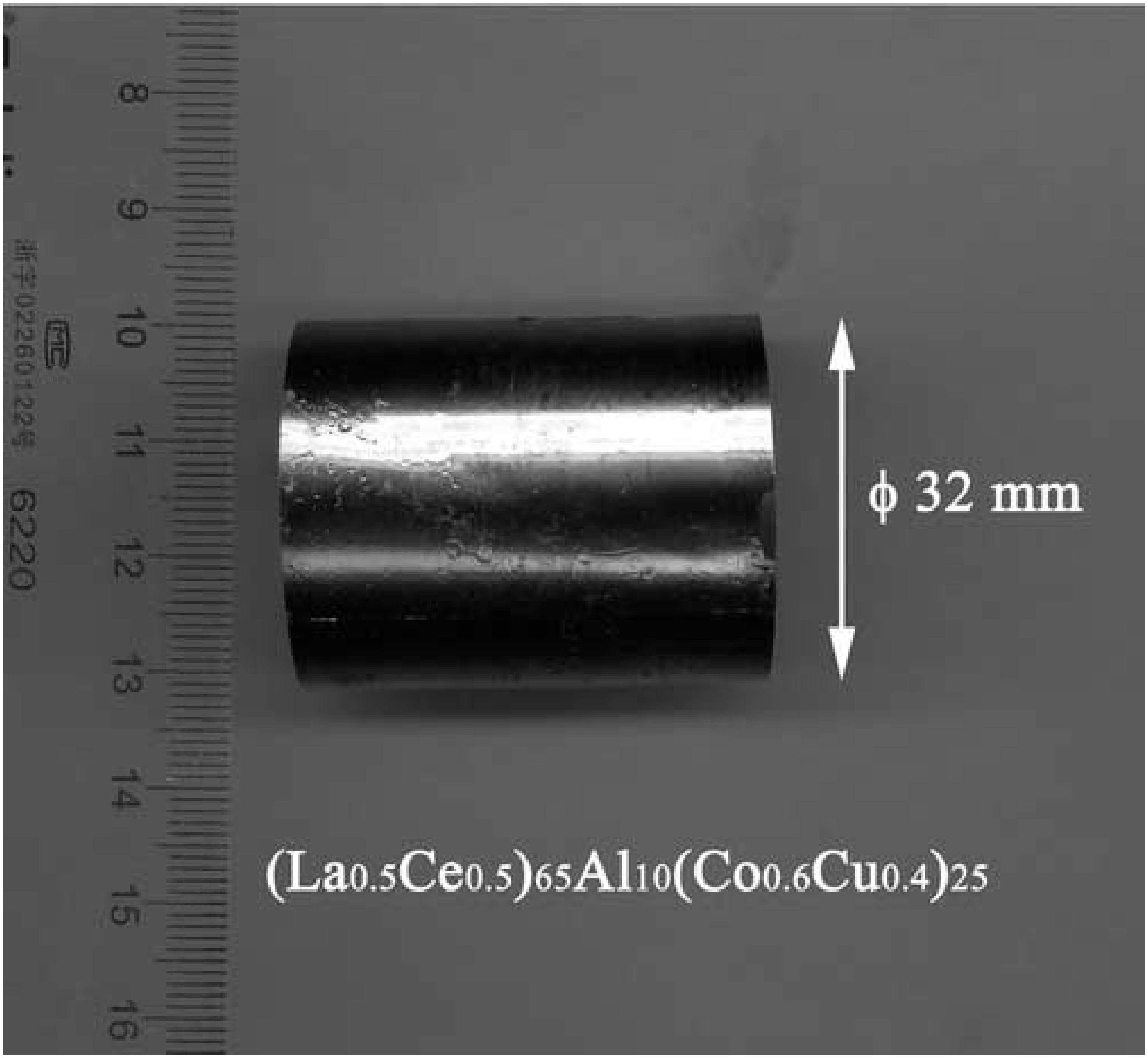

Fig. 2 XRD patterns of the as-cast rod of 32, 12 and 8 mm

Click here to download high resolution image

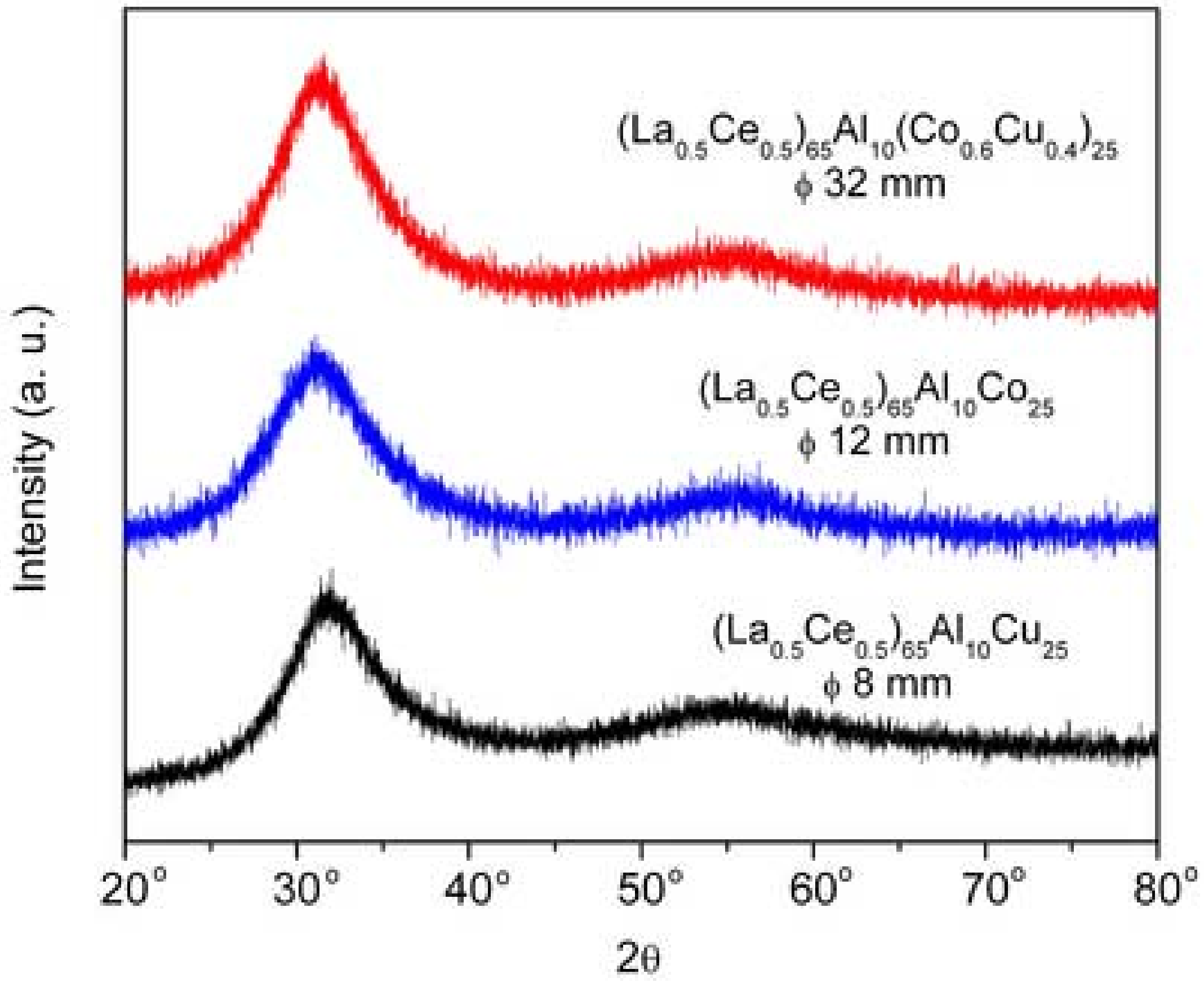

**Fig. 3 DSC curves of (La0.5Ce0.5)65Al10(Co0.6Cu0.4)25**

**Click here to download high resolution image**

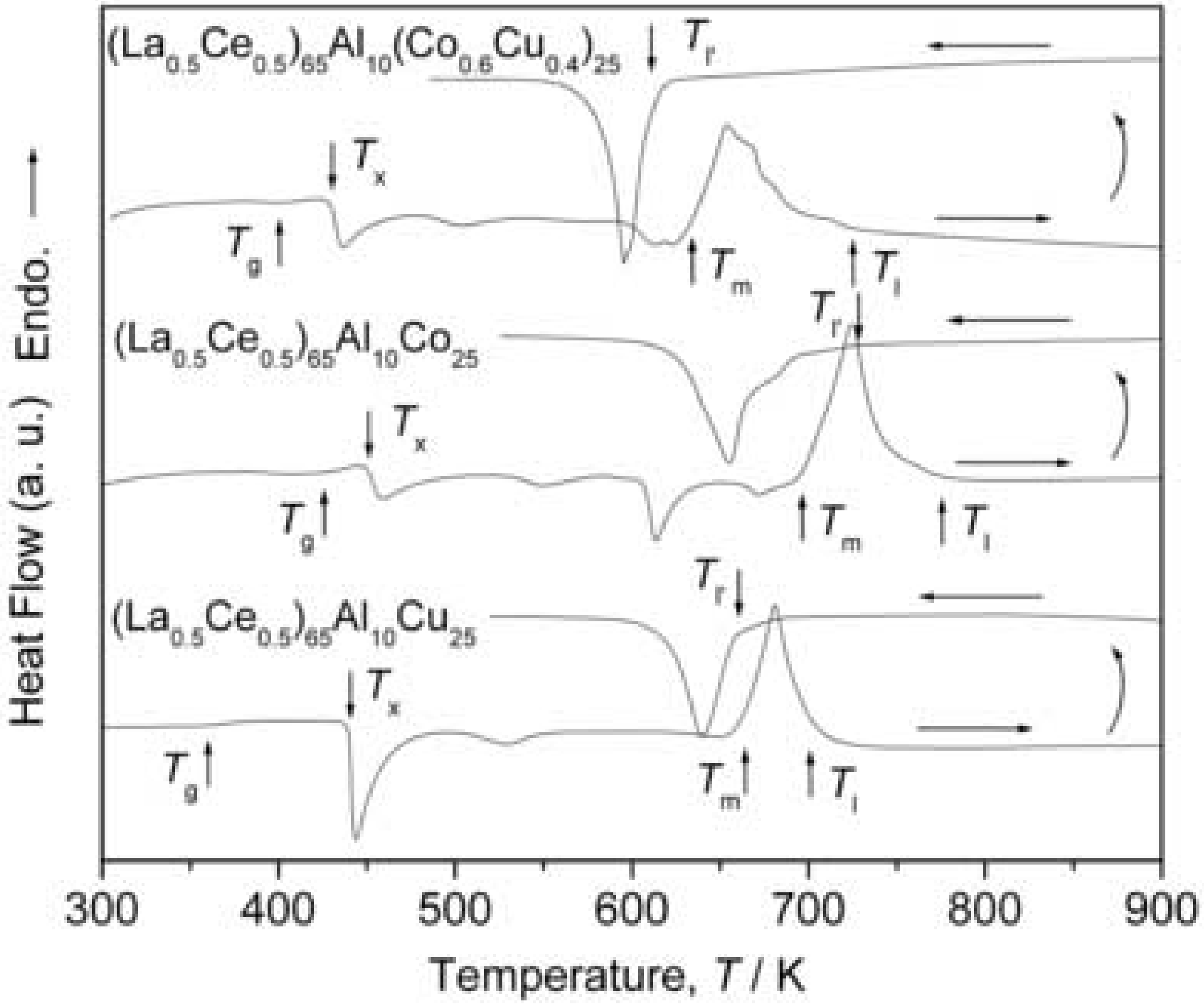

Fig. 4 DSC curves of the ternary Ln65Al10Co25

Click here to download high resolution image

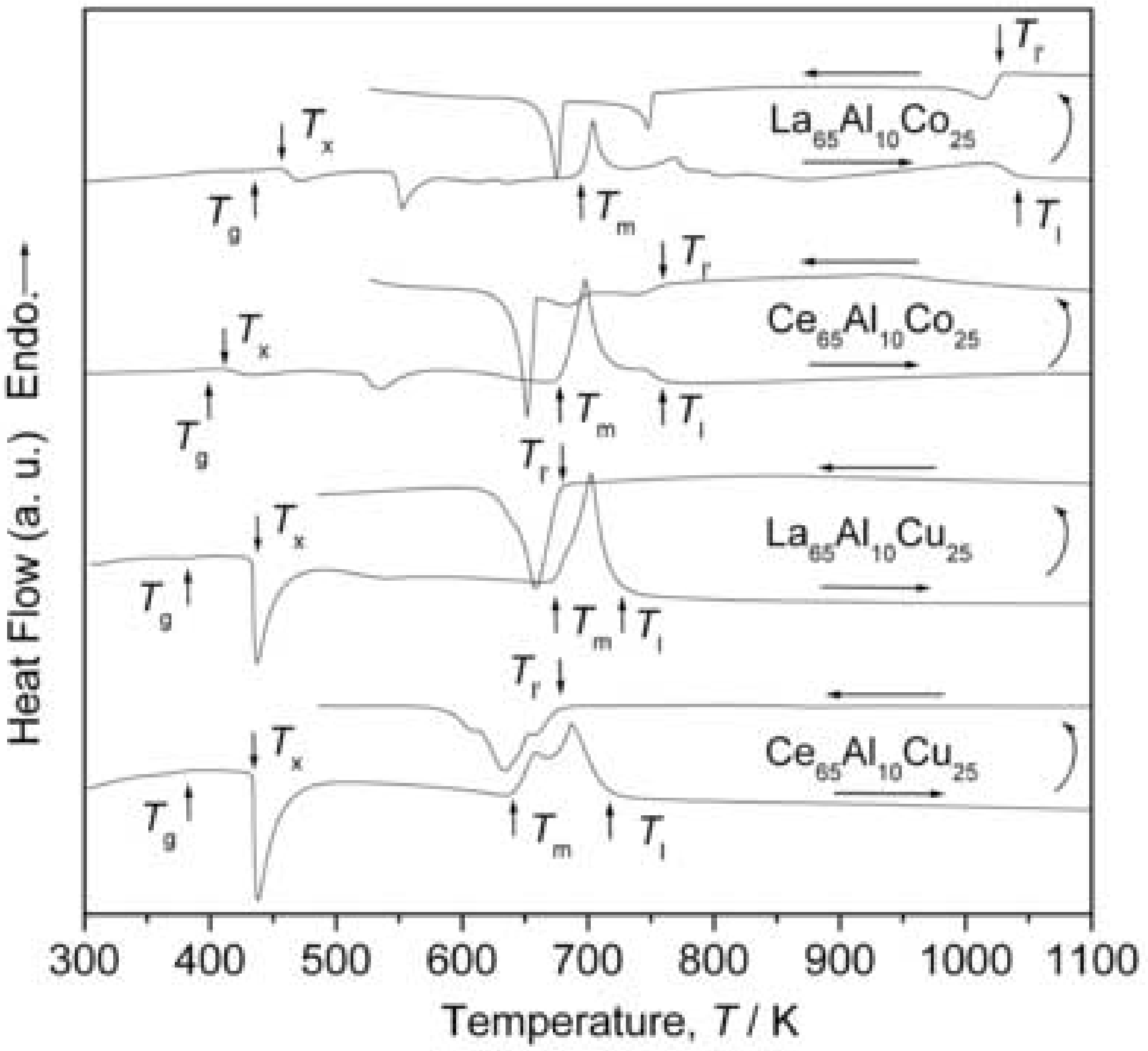

Table I Thermal parameters and critical diameters ($d_c$) of La-, Ce- and (La-Ce)-based glassy alloys.

| | $d_c$(mm) | $T_g$(K) | $T_x$(K) | $T_m$(K) | $T_l$(K) | $T_{l'}$(K) | $\Delta T_x$(K) | $T_g/T_m$ | $T_g/T_l$ | $T_g/T_{l'}$ | $T_x/(T_g+T_l)$ | $\Delta T_l$(K) |
|---|---|---|---|---|---|---|---|---|---|---|---|---|
| $La_{65}Al_{10}Co_{25}$ | 2 | 434 | 457 | 694 | 1041 | 1028 | 23 | 0.62 | 0.42 | 0.42 | 0.310 | 13 |
| $Ce_{65}Al_{10}Co_{25}$ | 2 | 396 | 410 | 676 | 758 | 755 | 14 | 0.58 | 0.52 | 0.52 | 0.355 | 3 |
| $La_{65}Al_{10}Cu_{25}$ | 4 | 382 | 432 | 673 | 722 | 680 | 50 | 0.57 | 0.53 | 0.56 | 0.391 | 42 |
| $Ce_{65}Al_{10}Cu_{25}$ | 4 | 367 | 433 | 641 | 715 | 675 | 66 | 0.57 | 0.51 | 0.54 | 0.400 | 40 |
| $(La_{0.5}Ce_{0.5})_{65}Al_{10}Cu_{25}$ | 8 | 360 | 440 | 662 | 701 | 660 | 80 | 0.54 | 0.51 | 0.54 | 0.415 | 41 |
| $(La_{0.5}Ce_{0.5})_{65}Al_{10}Co_{25}$ | 12 | 425 | 450 | 695 | 775 | 729 | 25 | 0.61 | 0.55 | 0.58 | 0.375 | 46 |
| $(La_{0.5}Ce_{0.5})_{65}Al_{10}(Co_{15}Cu_{10})$ | 32 | 400 | 430 | 634 | 724 | 610 | 30 | 0.63 | 0.55 | 0.66 | 0.382 | 114 |